\documentclass[sigplan]{acmart}
\AtBeginDocument{%
  }

\setcopyright{none}
\renewcommand\footnotetextcopyrightpermission[1]{}
\usepackage[english]{babel}
\usepackage{graphicx}
\usepackage{amsmath}
\usepackage{array}
\usepackage{parskip}
\usepackage{amsfonts}
\usepackage{stmaryrd}
\usepackage{microtype}
\usepackage{xspace}
\usepackage{hyperref}
\usepackage{thmtools}
\usepackage[capitalise]{cleveref}
\usepackage{tocloft}
\usepackage{csquotes}
\usepackage[all]{xy}     
\usepackage[vskip=3pt]{quoting}
\usepackage{etoolbox}

\usepackage{listings}
\newcommand{\BigSpace}{\hskip 1,5em}
\newcommand{\HugeSpace}{\hskip 3,5em}
\newcommand{\refToRule}[1]{\textsc{\small (#1)}}

\newcommand{\emath}[1]{\ensuremath{#1}\xspace}
\newcommand{\ple}[1]{\emath{{\langle #1 \rangle}}}

\newcommand{\fun}[3]{\emath{#1{:}\,#2 \rightarrow #3}} 
\newcommand{\funtype}[2]{\emath{#1\rightarrow #2}} 
\newcommand{\pfun}[3]{\emath{#1{:} #2 \rightharpoonup   #3}}

\newcommand{\N}{\mathbb{N}} 
\newcommand{\Subst}[3]   {#1[#2/#3]}

\newcommand{\NamedRule}[4]{{\scriptstyle{\textsc{(#1)}}}\
\displaystyle                  
\frac{#2}{#3}         
\begin{array}{l}
#4     
\end{array}
}

\newcommand{\NamedRuleOL}[3]{\scriptstyle{\textsc{(#1)}}\
\displaystyle  
#2\hfill#3
}

\newenvironment{grammatica}{\begin{array}{lcll}}{\end{array}}
\newcommand{\produzione}[3]{#1&::=&#2&\mbox{#3}}

\newcommand{\kw}[1]{\texttt{#1}}
\newcommand{\aux}[1]{\mathsf{#1}}
\newcommand{\mvar}[1]{\mathit{#1}}

\newcommand{\seq}[1]{\overline{#1}} 

\newcommand{\mnd}{\Mnd\mfun}
\newcommand{\mfun}{M}
\newcommand{\mun}{\eta}
\newcommand{\mmul}{\mu}

\newcommand{\mbind}{\mathbin{\gg=}} 
\newcommand{\Mmap}[3][]{\aux{map}\ifblank{#1}{}{_{#1}}\, #2\, #3}
\newcommand{\Mnd}[1]{\mathbb{#1}}

\newcommand{\ExSet}{\aux{Exc}}
\newcommand{\ExceptFun}[1][]{E\ifblank{#1}{}{_{#1}}}
\newcommand{\exc}{\aux{e}}

\newcommand{\PowerFun}{P}
\newcommand{\ListFun}{L} 
\newcommand{\elist}{\epsilon}
\newcommand{\cons}{\colon}

\newcommand{\lang}{\mathcal{L}}
\newcommand{\Exp}{\aux{Exp}}

\newcommand{\Val}{\aux{Val}} 
\newcommand{\val}{v} 
\newcommand{\Res}{\aux{Res}}

\newcommand{\Conf}{\aux{Conf}}
 
\newcommand{\valtoexp}{\aux{ret}} 
\newcommand{\Wrng}{\aux{Wr}}
 
\newcommand{\mexpr}{\textsc{e}} 
\newcommand{\mval}{\textsc{v}}

\newcommand{\finsem}[1]{\infsem[\star]{#1}}  
\newcommand{\infsem}[2][\infty]{\llbracket #2 \rrbracket_{#1}} 
\newcommand{\ehole}{[\ ]} 
\newcommand{\red}{\to}
\newcommand{\purered}{\red_p} 
\newcommand{\mrun}[1]{\emath{\aux{run}_{#1}}}

\newcommand{\lambdaEff}{\Lambda_{\Sigma}}
\newcommand{\f}{\mvar{f}}
\newcommand{\x}{\mvar{x}}
\newcommand{\y}{\mvar{y}}

\newcommand{\ve}{\mvar{v}} 
\newcommand{\e}{\mvar{e}}
\newcommand{\opg}{\mvar{op}} 

\newcommand{\Fun}[2]{\lambda #1.#2} 
\newcommand{\RecFun}[3]{\kw{rec}\,#1.\Fun{#2}{#3}}
\newcommand{\App}[2]{#1\,#2} 
\newcommand{\Ret}[1]{\kw{return}\ #1}

\newcommand{\Do}[3]{\kw{do}\ #1 = #2\kw{;}\ #3} 

\newcommand{\T}{\mvar{T}}
\newcommand{\TVar}[2]{#2:#1}
\newcommand{\TVars}[2]{\seq#2{:}\seq#1}
\newcommand{\TEff}[2]{#1{!}#2}
\newcommand{\FunType}[3]{#1{\rightarrow}_{#2}#3}
\newcommand{\eff}{\mvar{E}} 
\newcommand{\meet}{\wedge} 
\newcommand{\join}{\vee}        

\newcommand{\EffSet}{\aux{Eff}}

\newcommand{\EComp}[2]{#1{\cdot}#2}

\newcommand{\SetOf}[1]{\{#1\}}

\newcommand{\IsWFExp}[3]{#1\vdash#2:#3}
\newcommand{\IsWFGroundExp}[2]{\vdash#1:#2}
\newcommand{\IsEffExp}[4]{#1\vdash#2:\TEff{#3}{#4}}
\newcommand{\IsEffGroundExp}[3]{\vdash#1:\TEff{#2}{#3}}
\newcommand{\subt}{\leq}
\newcommand{\sube}{\eord}
\newcommand{\SubT}[2]{#1\subt#2}
\newcommand{\SubTE}[2]{#1\subt#2}
\newcommand{\SubE}[2]{#1\sube#2}

\newcommand{\SubEI}[2]{#1\subseteq#2}

\newcommand{\TS}{\Theta} 
\newcommand{\Types}{\aux{Ty}}
\newcommand{\ty}{\tau}
\newcommand{\MEff}{\mathcal{E}}
\newcommand{\Eff}{\aux{Eff}}
\newcommand{\ef}{\varepsilon}
\newcommand{\eord}{\preceq}
\newcommand{\emul}{\cdot}
\newcommand{\eun}{1} 

\newcommand{\VWT}[1][]{\aux{WT}^{\aux{V}}\ifblank{#1}{}{_{#1}}}
\newcommand{\EWT}[1][]{\aux{WT}^{\aux{E}}\ifblank{#1}{}{_{#1}}}

\newcommand{\X}{\mvar{X}}
\newcommand{\Y}{\mvar{Y}}
\newcommand{\ebasic}{\aux{basic}}

\newcommand{\Nat}{\texttt{Nat}}
\newcommand{\Bool}{\texttt{Bool}}

\newcommand{\If}[3]{\kw{if}\ #1\ \kw{then}\ #2\ \kw{else}\ #3}
\newcommand{\Num}{\texttt{n}}
\newcommand{\True}{\texttt{true}}
\newcommand{\False}{\texttt{false}}

\newcommand{\withindex}[2]{#1\langle#2\rangle}
 \newcommand{\raiseop}[1]{\withindex{\tt raise}{#1}}

\newcommand{\chooseop}{{\tt choose}}

\newcommand{\hc}{\mvar{c}}
\newcommand{\handler}{\mvar{h}}
\newcommand{\Handler}[3]{#1, #2\mapsto#3} 
\newcommand{\With}[2]{\kw{handle}\ #2\ \kw{with}\  #1}
\newcommand{\WithLong}[4]{\kw{handle}\ #4\ \kw{with}\  \Handler{#1}{#2}{#3}}

\newcommand{\HC}[4]{#1(#2)\mapsto_{#4}#3} 
\newcommand{\mode}{\mu}
\newcommand{\Continue}{\aux{c}}
\newcommand{\Stop}{\aux{s}}
\newcommand{\cfilter}{\mvar{C}}
\newcommand{\CFilter}[3]{#1\mapsto_{#2}#3}
\newcommand{\hfilter}{\mvar{H}}
\newcommand{\HFilter}[2]{#1, #2} 
\newcommand{\FilterF}[1]{\mathcal{F}_{#1}}
\newcommand{\FilterFun}[2]{\FilterF{#2}(#1)}
\newcommand{\FilterFEx}[1]{\widehat{\mathcal{F}}_{#1}}

\newcommand{\IsWFHandler}[5]{#1;#2\vdash#3:\TEff{#4}{#5}}
\newcommand{\IsWFClause}[4]{#1;#2\vdash#3:#4}

\newcommand{\ESOP}[2]{Programming Languages and Systems - #1 European Symposium on Programming, {ESOP} #2} 
\newcommand{\FOSSACS}[2]{Foundations of Software Science and Computation Structures, #1 International Conference, {FoSSaCS} #2}

\newcommand{\LICS}[2]{Proceedings of the #1 Annual {ACM/IEEE} Symposium on Logic in Computer Science, {LICS} #2}
\newcommand{\MFPS}[2]{The #1 Conference on Mathematical Foundations of Programming Semantics, {MFPS} #2} 
\newcommand{\POPL}[2]{Proceedings of the #1 {ACM/SIGPLAN} Symposium on Principles of Programming Languages, {POPL} #2}

\begin{document}
\title{A monadic interpreter and type-and-effect checker}


\author{Stefano Raviola}
\affiliation{%
  \institution{DISIT, Universit\`a del Piemonte Orientale}
  \city{Alessandria}
  \country{Italy}}
\email{20051799@studenti.uniupo.it}

\author{Paola Giannini}
\affiliation{%
  \institution{DiSSTE, Universit\`a del Piemonte Orientale}
  \city{Vercelli}
  \country{Italy}}
\email{paola.giannini@uniupo.it}
  
\author{Francesco Dagnino}
\affiliation{%
  \institution{DIBRIS, Universit\`a di Genova}
  \city{Genova}
  \country{Italy}}
\email{francesco.dagnino@dibris.unige.it}

\renewcommand{\shortauthors}{S. Raviola, P. Giannini, F. Dagnino}

\begin{abstract}

We present a concrete implementation in Haskell of a monadic framework that includes both a small-step interpreter and a type-and-effect checker for the corresponding language. Our approach separates the language syntax from the semantics of its effects. This design allows the interpreter to remain parametric over the underlying monad, while the static checker approximates effects independently of their concrete implementation.

The theoretical foundation of this framework—applied to a call-by-value lambda calculus with generic effects represented by operations that produce monadic values and are managed through handlers—was introduced in previous work, where the focus was on proving the soundness of the approach. In contrast, the present work leverages Haskell’s support for modular programming and monads to demonstrate that the framework is practically implementable and usable.

We illustrate the approach with examples using the monad of exceptions and the one of nondeterminism and expressions both with and without handlers.

\end{abstract}

\maketitle



\section{Introduction}
\label{sec:introduction}

Moggi’s work \cite{Moggi89,Moggi91} established monads as a modular foundation for the denotational semantics of effectful languages, by distinguishing \emph{pure} expressions from \emph{monadic} (effectful) computations. Haskell first\footnote{Later followed by many other languages, including Scheme, Python, Racket, Scala, and F$\#$.} showed that this approach scales to mainstream programming via a monadic type constructor encapsulating effects. However, monads\footnote{In Haskell, the methods of the \lstinline{Monad} typeclass.} do not intrinsically provide operations for raising effects, which must be defined ad hoc. \emph{Algebraic and generic effects} \cite{PlotkinP01,PlotkinP02,PlotkinP03} address this limitation by making effect operations explicit and interpreting them through additional monadic structure. This approach, combined with handlers \cite{PlotkinPretnar09,PlotkinPretnar13,BauerP15,Pretnar15}, has been exploited in programming languages, e.g., in Scala and \mbox{OCaml 5.} 

Operational semantics describes how a program is executed. It typically comes in two forms: big-step semantics, which reduces expressions directly to their final values, and small-step semantics, which describes computation as a sequence of atomic reduction steps. While big-step semantics is often simpler for defining interpreters, small-step semantics is the preferred choice when establishing properties such as type soundness,  see \citet{WrightF94}. 

 In \citet{DGZ25},  a novel approach to statically reason not only about results but also about the effects that computations may produce is proposed, which builds on the two foundational ideas in programming language theory: the use of monads to model computational effects, as introduced by Moggi, and the operational approach to type soundness based on progress and subject reduction. The type system is extended to a type-and-effect system, and an operational meta-theory of monadic type-and-effect soundness is developed, analogous to the classical small-step framework for type soundness.  Program expressions reduce deterministically to monadic expressions, and this reduction is extended to a total function so that reduction steps compose via Kleisli composition. As a result, computations yield infinite reduction sequences, with termination represented by special monadic results that do not raise further effects.

In this paper, we describe a Haskell implementation of a type-and-effect checker and a small-step interpreter for an extension of the lambda calculus with generic effects and effect handlers  of~\citet{DGZ25}.  The modules separate the generic framework from the  specific language and algebraic effects. Examples of tracking exceptions and controlling nondeterminism are implemented. The tool provides a front-end that can be used to specify the monad for interpreting the effects raised by the execution of the effectful operations and the expression to be evaluated. 

The implementation has the following properties.
\begin{itemize}
\item The reduction engine is \emph{parametric over the underlying monad}. 
  It executes the ``effectful operations''  by delegating their execution to an abstract interface, so that the runtime behavior can change, e.g., from exceptions to non-determinism, without modifying the core rules.
\item The \emph{type-and-effect checker} infers both types and effects and checks that they are compatible with the ones specified in the expression. The abstract effect types are instantiated as sets of effectful operations. 
\end{itemize}
The implementation aims to show that the theoretical framework can be used to provide a practical embedding of generic effects in a programming language. To achieve this, the language introduced  by~\citet{DGZ25}  is extended with a conditional construct. This extension requires the addition of a subtyping relation on base types, as well as the inclusion of meet and join operators in the algebraic structure of effect types.

\paragraph{Outline of the paper} In \cref{sec:background} we present the definitions underlying the monadic interpreter and type-and-effect checker. $\lambdaEff$ with its operational semantics and type system are introduced in \cref{sec:language}. \cref{sec:implementation} contains a description of the Haskell implementation with an outline of its modules, and two examples of the small-step monadic reduction are given in \cref{sec:examples}. In \cref{sec:related} we compare with other proposals of similar work and finally \cref{sec:conclusion} draws some conclusions and outlines future work.

\Cref{sec:background,sec:language} present material largely drawn from \citet{DGZ25}. The language and type-and-effect system described in \cref{sec:language} extend that of \cite{DGZ25} by introducing a conditional construct, meet and join operators on types and a richer algebraic structure for effect types. Moreover, the abstract effect type structure is instantiated on sets of operations.  The remaining sections are original.

 
 \section{Background}
\label{sec:background}
In this section we introduce the theoretical background from \cite{DGZ25} needed in the paper.

{\em Monads} \cite{EilenbergM65,Street72} are a fundamental notion in category theory, enabling an abstract and unified study of algebraic structures. 
Since Moggi's seminal papers \cite{Moggi89,Moggi91}, they have also  become a major tool in computer science, especially for describing the semantics of computational effects, and integrating them into programming languages in a structured and principled way. 
In functional programming, monads $\mnd = \ple{\mfun,\mun,\mbind}$ are defined by a mapping $\mfun$ from pure values to monadic values and two operations:
\begin{itemize}
\item $\mun: \X \to \mfun \X$, which lifts a value of type $\X$ to a monadic value of type $\mfun \X$ and
\item $\mbind: \mfun \X \to (\X \to \mfun \Y) \to \mfun \Y$
\end{itemize}
The operator $\mbind$, also called $\aux{bind}$, corresponds, intuitively, to the sequential composition of two expressions with effects, where the latter  depends on a parameter \emph{bound} to the result of the former. 

The monad of {\em exceptions} can be used to signal an anomaly in the execution of a program.
\begin{example}[Exceptions] \label{ex:exception-mnd} 
Let us fix a set $\ExSet$ of exceptions. 
The monad $\ExceptFun[\ExSet]$ is given by 
$\ExceptFun[\ExSet] X = \ExSet + X$, and 
\begin{quoting}
\begin{math}
\begin{array}{c}
\mun(x) = \iota_2(x) 
\\[1em]
\alpha \mbind f = \begin{cases}
f(x)\ \mbox{if}\ \alpha = \iota_2(x) \\
\alpha\ \text{otherwise ($\alpha = \iota_1(\exc)$ for some $\exc\in \ExSet$)} 
\end{cases}  
\end{array}
\end{math} 
\end{quoting}
where $+$ denotes disjoint union (coproduct) and $\iota_1,\iota_2$ are the left and right injections, respectively. 
We will omit the reference to the set $\ExSet$ when it is clear from the context. 
\end{example}

The monad of {\em non-determinism}  is used to model nondeterministic computations which may return an arbitrary number of results.
\begin{example}[Classical Non-Determinism]\label{ex:pow-mnd}
The monad ${\PowerFun}$ is given by 
$\PowerFun \X = \wp(\X)$, that is, $\PowerFun \X$ is the set of all subsets of $\X$, and 
\begin{quoting}
\begin{math}
\mun(x) =\{x\} \HugeSpace
\alpha \mbind f  = \bigcup_{x\in\alpha} f(x) 
\end{math} 
\end{quoting}
A variant of this monad is 
the {\em list} monad $\ListFun$, where 
the set $\ListFun \X$ of (possibly infinite) lists over $\X$ is coinductively defined by: 
$\elist\in\ListFun(\X)$ and, 
if $x\in \X$ and $l\in\ListFun(\X)$, then $x\cons l \in\ListFun(\X)$. 
We use the notation $[x_1,\ldots,x_n]$ to denote the finite list $x_1\cons \ldots\cons x_n\cons\elist$. 
Here $\mun(x) = [x]$ and the monadic bind is corecursively defined by the following clauses: 
$\elist\mbind f = \elist$ and 
$(x\cons l)\mbind f = f(x)(l\mbind f)$, 
where juxtaposition denotes the concatenation of possibly infinite lists. 
 $\ListFun$ is the monad provided by Haskell which we used in our implementation of non-determinsm. 
\end{example} 

The {\em monadic operational semantics} for a language $\lang$ is characterized by the following components.
\begin{definition}\label{def:mnd-sem} 
Let $\lang$ be a triple $\ple{\Exp,\Val,\valtoexp}$, called a \emph{language}, with $\Exp$ the set of \emph{expressions}, $\Val$ the set of \emph{values}, and $\fun{\valtoexp}{\Val}{\Exp}$ an injective function. 
A \emph{monadic operational semantics} for $\lang$ consists of:
\begin{itemize}
\item a monad $\mnd$
\item  a  relation $\red \subseteq \Exp\times\mfun\Exp$, called \emph{monadic (one-step) reduction},  such that 
\begin{itemize}
\item $\red$ is a partial function and 
\item for all $\val\in\Val$, $\valtoexp(\val)\not\red$. 
\end{itemize}
\end{itemize}
\end{definition} 
The set $\Exp$  contains  expressions that can be executed, while $\Val$ contains values produced by the computation. 
The inclusion $\valtoexp$ identifies the expressions representing successful termination with a given value. 
The elements of $\mfun\Exp$, called \emph{monadic expressions}, are the counterpart of expressions in the monad $\mnd$. 
The relation $\red$ models single computation steps, which transform expressions into monadic ones, thus possibly raising computational effects. 
Finally, the first requirement on $\red$ ensures that it is deterministic, while the latter ensures that expressions representing values cannot be reduced. 

Starting from $\red \subseteq \Exp\times\mfun\Exp$ a reduction between monadic configurations can be defined as shown at the beginning of \cref{sec:modules}.

\emph{Algebraic and generic effects} were introduced by Plotkin and Power,  \cite{PlotkinP01,PlotkinP02,PlotkinP03}. They are essentially
a set of {\em operations} that represent the sources of effects. These effect operations are added as additional monadic structure. In our setting
we assume, for each operation $\opg$ with arity $k$, a partial function ${\pfun{\mrun{\opg}}{\Val^k}{\mfun\Val}}$,  returning a monadic 
value expressing the effects raised by a call of the operation. 
The function could be undefined, for instance when arguments do not have the expected types. 

 \begin{example}\label{ex:genericEffect}
 \begin{enumerate}
 \item \label{ex:genericEffect:ex} Take the monad of \cref{ex:exception-mnd}. For each $\exc\in\ExSet$, we assume an operation  $\raiseop{\exc}$, with
\begin{quoting}
$\fun{\mrun{\raiseop{\exc}}}{\mathbf{1}}{\mfun{\Val}}$\HugeSpace
$\mrun{\raiseop{\exc}}=\iota_2(\exc)$
\end{quoting}
 \item \label{ex:genericEffect:nd} Take the monad of \cref{ex:pow-mnd}  in the variant  of the possibly infinite lists. We assume a constant operation $\chooseop$,  with 
\begin{quoting}
$\fun{\mrun{\chooseop}}{\mathbf{1}}{\mfun{\Val}}$\HugeSpace
$\mrun{\chooseop}=[\True,\False]$
\end{quoting}
 \end{enumerate}
 \end{example}

The last ingredient  of the framework is a {\em type-and-effect system}.  We can abstract a  type-and-effect system with a predicate 
over expressions indexed not only by types but also by \emph{effect types}, describing the computational effects that  expressions 
can produce during their evaluation, as defined below. 

\begin{definition}\label{def:mnd-type-system} 
Let $\lang=\ple{\Exp,\Val,\valtoexp}$ be a language.
A \emph{type-and-effect system} 
$\TS = \ple{\Types,\MEff,\EWT,\VWT}$ for $\lang$  consists of the following data: 
\begin{itemize}
\item a set $\Types$ of \emph{types}
\item an ordered monoid $\MEff = \ple{\Eff,\eord,\emul,\eun}$ of \emph{effect types}
\item for every $\ty\in\Types$ and $\ef\in\Eff$, predicates 
$\VWT[\ty]\subseteq\Val$ and $\EWT[\ty,\ef]\subseteq\Exp$  such that 
\begin{itemize}
\item $\ef\eord\ef'$ implies $\EWT[\ty,\ef]\subseteq\EWT[\ty,\ef']$ and 
\item $\valtoexp(\val) \in \EWT[\ty,\ef]$ iff $\val\in\VWT[\ty]$ and $\eun\eord\ef$
\end{itemize}
\end{itemize}
\end{definition}
The ordered monoid is a typical structure for effect systems \cite{NielsonN99,MarinoM09,Katsumata14}:
$\eun$ represents the absence of computational effects, 
$\ef_1\emul\ef_2$ represents the composition of computational effects described by $\ef_1$ and $\ef_2$, and 
$\ef_1\eord\ef_2$ states that the effect type $\ef_1$ is \mbox{more specific than $\ef_2$. }

The two families $\VWT$ and $\EWT$ are, for each index, predicates over values and expressions, respectively: 
$\VWT[\ty]$ is the set of values of type $\ty$, and 
$\EWT[\ty,\ef]$ is the set of expressions of type $\ty$ which may raise effects described by $\ef$. 
The first requirement, that is, monotonicity with respect to the order, states that the latter actually models the fact that if $\ef_1\eord\ef_2$, then $\ef_1$ is really more specific than $\ef_2$. 
The second requirement states that an expression which is the embedding of a value has the same type,  and an effect type which is not forcing any effect.

\section{$\lambdaEff$: A Lambda Calculus with Generic Effects}\label{sec:language}

In this section we present the core of $\lambdaEff$, the language for which we implemented the interpreter and type-checker. 
The language implemented also has natural number and boolean operators with the standard operational semantics and typing.

\subsection{Syntax}
The abstract syntax of $\lambdaEff$ is shown in \cref{fig:syntax}. 
\begin{figure}[th]
\[
\begin{array}{lll}
\ve &::= \Num \mid \True \mid \False \mid \Fun{\x}{\e} \mid \RecFun{\f}{\x}{\e} \mid x & \text{value} \\
e &::= \App{\ve_1}{\ve_2} \mid \If{\ve}{\e_1}{\e_2}  \\
  &\quad\mid \opg(\seq\ve)\mid \With{\handler}{\e} \\
  &\quad\mid \Do\x{\e_1}{\e_2}  \mid \Ret\ve & \text{expression} \\
\handler &::= \Handler{\seq\hc}{\x}{\e} & \text{handler} \\
\hc &::=\HC{\opg}{\seq\x}{\e}{\mode} & \text{clause} \\
\mode &::= \Continue\mid\Stop & \text{mode}
\end{array}
\]
\caption{$\lambdaEff$: fine-grain syntax}\label{fig:syntax} 
\end{figure}

 We use $\seq{\ve}$ as a metavariable for sequences $\ve_1,\ldots, \ve_n$, 
and analogously for other sequences.
We assume variables such as $\x$, $\y$, $\f$, and so on, using the latter to range over function variables.
Because evaluation order matters in the presence of effects, we distinguish between inert {\em values}, $\ve$, and potentially effectful {\em expressions}, $\e$ (also called {\em computations}). This follows the fine-grain approach of \cite{LevyPT03}.

The language includes ordinary functions, $\Fun{\x}{\e}$, and recursive functions, $\RecFun{\f}{\x}{\e}$, where the function with parameter $\x$ and body $\e$ may refer to itself recursively through $\f$. Standard application and conditional constructs are provided.

Effects are triggered by invoking operations, $\opg(\seq\ve)$. Handlers serve as mechanisms for managing such effects, similar to exception handling in mainstream programming languages. In particular, $\With{\handler}{\e}$ consists of a {\em final expression} $\e$ together with a sequence of {\em clauses} $\seq c$, viewed as a map, so that each operation has at most one corresponding clause. When an operation is invoked, the matching clause (if present) determines how it is handled. Afterward, the final expression may or may not be evaluated, depending on the {\em mode}: either $\Continue$ or $\Stop$. A $\Continue$-clause replaces an effect with alternative behavior while preserving the flow of execution, whereas a $\Stop$-clause interrupts the normal computation once the effect is handled.

Sequential evaluation is expressed by $\Do\x{\e_1}{\e_2}$. Effects arising during the evaluation of $\e_1$ are propagated to $\e_2$, as made precise by the operational semantics. The construct $\Ret\ve$ embeds a value $\ve$ into the chosen monad.

We write $\Exp$ and $\Val$ for the sets of closed expressions and values of $\lambdaEff$, respectively. 

\subsection{Operational Semantics}

The (one-step) monadic reduction of the language, parameterized by a monad $\mnd = \ple{\mfun,\mun,\mmul}$, is given by a relation $\red \subseteq \Exp \times \mfun\Exp$. We use $\mval$ and $\mexpr$ to range over elements of $\mfun\Val$ and $\mfun\Exp$, respectively.

This reduction is defined on top of a {\em pure reduction} $\purered \subseteq \Exp \times \Exp$. In our setting, the pure reduction evaluates function applications by substituting into their bodies, reduces conditional expressions to the chosen branch, and processes handlers. All remaining expressions are in normal form with respect to pure reduction and are therefore evaluated with the rules of the {\em monadic reduction}.
 
\begin{figure}[th]
\begin{math}
\begin{array}{l}
\NamedRuleOL{app}
{ \App{\ve}{\ve'} \purered \Subst{\e}{\ve'}{\x}}
{\boxed{\ve =\Fun{\x}{\e}}}
\\[1.3em]
\NamedRuleOL{app-r}
{ \App{\ve}{\ve'} \purered \Subst{\Subst{\e}{\ve}{\f}}{\ve'}{\x}\quad\quad}
{\boxed{\ve =\RecFun{\f}{\x}{\e}}}\\[1.3em]
\NamedRuleOL{if-t}
{ \If{\True}{\e_1}{\e_2} \purered \e_1}
{}\\[1.3em]
\NamedRuleOL{if-f}
{ \If{\False}{\e_1}{\e_2} \purered \e_2}
{}\
\end{array}
\end{math}
\caption{Pure reduction: application and conditional}\label{fig:pure-red} 
\end{figure}

\begin{figure}[th]
\begin{math}
\begin{array}{l}
\text{let  }\boxed{\handler=\Handler{\seq\hc}{\x}{\e'}}\text{  in}
\\[2ex]
\NamedRuleOL{w-do}
{ \begin{array}{l}
\With{\handler}{\Do{\y}{\e_1}{\e_2}}\purered 
\end{array}
}
{
}\\
\quad\quad\quad\quad\quad\WithLong{\seq\hc}{\y}{(\With{\handler}{\e_2})}{\e_1}
\\[1.3ex]
\NamedRuleOL{w-ret}
{ \With{\handler}{\Ret\ve}\purered \Do\x{\Ret\ve}{\e'} }
{
 } 
\\[1.3ex] 
\NamedRuleOL{w-cont}
{ 
\With{\handler}{\opg(\seq\ve)} \purered \Do{\x}{\Subst{\e}{\seq\ve}{\seq\x}}{\e'}
}
{ 
}
\\ \hfill\boxed{\HC{\opg}{\seq\x}{\e}{\Continue}\in\seq\hc}
\\[1.3ex]
\NamedRuleOL{w-stop}
{ 
\With{\handler}{\opg(\seq\ve)}\purered \Subst{\e}{\seq\ve}{\seq\x}\ \
}
{ 
\boxed{\HC{\opg}{\seq\x}{\e}{\Stop}\in\seq\hc}
}
\\[1.3ex]
\NamedRuleOL{w-fwd}
{\With{\handler}{\opg(\seq\ve)} \purered \Do{\x}{\opg(\seq\ve)}{\e'} }\  \
{  
\boxed{{\opg}\not\in\seq\hc}
}
\\[1.3ex]
\NamedRule{w-ctx}{
  \e \purered \e'
}{ \With{\handler}{\e}  \purered \With{\handler}{\e'}}
{ }
\end{array}
\end{math}
\caption{Pure reduction with handlers}\label{fig:pure-red-handlers} 
\end{figure}
The rules for application and conditional expressions of \cref{fig:pure-red} are obvious. 

The semantics of a handled expression, see \cref{fig:pure-red-handlers},  depends on the syntactic form of the expression being handled.

If the handled expression is a \lstinline{do} composition of two subexpressions, the outer \lstinline{do} is eliminated by reducing to the first subexpression, while taking the second one as the final expression. The handler clauses are propagated to both subexpressions.
If the handled expression is a \lstinline{return}, the handler itself disappears, reducing to a \lstinline{do} composition of the returned value and the final expression.

When the handled expression is an operation call, the behavior depends on whether a corresponding clause exists. If a matching clause is found, its body is executed after substituting the formal parameters with the actual arguments (as described in the rules  \refToRule{w-cont} and \refToRule{w-stop}). In the case of a $\Continue$-clause, the final expression is evaluated afterward as well. If no matching clause is present, the handler is removed, reducing the term to a \lstinline{do} composition of the operation call and the final expression. This effectively forwards the operation outward, allowing it to be handled at an enclosing level. The contextual rule follows the standard pattern.


The rules for monadic reduction are presented in \cref{fig:monadic-red}. As noted earlier, they are parameterized by the chosen monad and rely on the following components:
\begin{itemize}
\item A function $\fun{\mun_\Exp}{\Exp}{\mfun\Exp}$ that embeds expressions into the monad. In this section, we simply write this embedding as $\mun$.
\item  A function $\fun{\aux{map}}{(\funtype{\Exp}{\Exp})}{\funtype{\mfun\Exp}{\mfun\Exp}}$ that lifts functions on expressions to functions on monadic expressions.
\end{itemize}
In addition, as already mentioned, for each operation $\opg$ of arity $k$, we assume a partial function $\pfun{\mrun{\opg}}{\Val^k}{\mfun\Val}$, which yields a monadic value representing the effects produced by invoking the operation. 

\begin{figure}[th]
\begin{math}
\begin{array}{l}
\NamedRule{pure}{
  \e \purered \e' 
}{ \e \red \mun(\e') }
{ } 
\\[5ex]
\NamedRuleOL{effect}
{ \opg(\seq\ve) \red \Mmap{(\Ret{\ehole})}{\mrun{\opg}(\seq\ve)}  }
{ 
}
\\[5ex]
\NamedRuleOL{ret}
{ \Do{\x}{\Ret\ve}{\e} \red \mun(\Subst\e\ve\x) }
{ } 
\\[5ex] 
\NamedRule{do}{
  \e_1 \red \mexpr 
}{ \Do\x{\e_1}{\e_2} \red \Mmap{(\Do\x{\ehole}{\e_2})}{\mexpr} } 
{ }
\end{array}
\end{math}
\caption{Monadic (one-step) reduction}\label{fig:monadic-red} 
\end{figure}
Rule \refToRule{pure}  lifts a pure reduction step into the monad by embedding its result.
Rule \refToRule{effect} is responsible for actually raising an effect. It applies the function of type $\funtype{\mfun\Val}{\mfun\Exp}$ obtained by lifting, via $\aux{map}$, the context $\Ret{\ehole}$ to the monadic value resulting from the operation call. Here, the context $\Ret{\ehole}$ (an expression with a hole) is identified with the function  $\ve \mapsto \Ret\ve$  of type $\Val \to \Exp$.

In rule \refToRule{ret}, when the first subexpression of a \lstinline{do} construct evaluates to a value, the term reduces to the monadic embedding of the second subexpression, after substituting the bound variable with that value.
Rule \refToRule{do}, by contrast, propagates the reduction of the first subexpression.  In this rule $\mexpr$ is a metavariable denoting a monadic expression.  To account for possible effects, we lift, again via $\aux{map}$, the context $\Do\x{\ehole}{\e_2}$ to obtain a function of type $\funtype{\mfun\Exp}{\mfun\Exp}$, which is then applied to the monadic result of $\e_1$. As before, the context $\Do\x{\ehole}{\e_2}$ is identified with the function  $\e \mapsto \Do{\x}{\e}{\e_2}$  of type $\Exp \to \Exp$.

\subsection{Type-and-Effect System}
We present the {\em type-and-effect system} for $\lambdaEff$ needed for \cref{def:mnd-type-system}.  We introduce value and 
effect types, $\T$ and $\eff$, with their subtyping relation and   present the typing rules for the judgments $\IsWFExp{\Gamma}{\ve}{\T}$ and  $\IsEffExp{\Gamma}{\exp}{\T}{\eff}$. Then, we define the sets $\VWT[\ty]$ and $\EWT[\ty,\ef]$. 

\begin{figure}[th]
\begin{math}
\begin{grammatica}
\produzione{\T}{\Nat\mid\Bool\mid\FunType{\T}{\eff}{\T'}\mid\top\mid\bot}{type}\\
\produzione{\Gamma}{\seq{\x:\T}}{context}
\end{grammatica}
\end{math}
\caption{Types and contexts}\label{fig:typed}

\end{figure}
Types are defined in \cref{fig:typed}. We have primitive types and functional types, as in \cite{DGZ25}.  In addition we assume to have 
a bottom and top type.  \emph{Effect types} (\emph{effects} when there is no ambiguity), ranged over by $\eff$ are meant to be static approximations of the computational effects raised by an expression. Functional types are decorated by effects that over-approximate the effects that may be raised by the evaluation of the body of the function.

In our implementation, to type conditional expressions  and operations calls , we require, in addition to the monoidal structure of \cref{def:mnd-type-system}, the {\em meet}, $\meet$, and {\em join}, $\join$, operators,  as well as the operation injection and effect filtering. That is  \\
\centerline{$\MEff = \ple{\Eff,\eord,\meet, \join,\emul,\eun ,\ebasic, \FilterF{\cdot}}$.} 
We also provide a concrete instantiation for $\MEff$ that models an effect type 
as a {\em set of operations} $\eff \subseteq \Sigma$. Therefore
\begin{itemize}
\item subeffecting corresponds to set inclusion:
    $\eff_1 \preceq \eff_2$ if $\eff_1\subseteq\eff_2 $;
\item meet and join correspond to set intersection and union:
    $\eff_1 \meet  \eff_2 = \eff_1 \cap \eff_2$ and $\eff_1 \join  \eff_2 = \eff_1 \cup \eff_2$;
\item $\emul$  is set union, $\eff_1 \emul \eff_2 = \eff_1 \cup \eff_2 $,  with unit the empty set of operations, $\eun=\emptyset$;
\item  $\ebasic$ is the singleton set $\SetOf{op}$, for every $\opg \in \Sigma$; 
\item  $\FilterF{\cdot}$ is the filter function defined  below, see (1).  
\end{itemize}
In future work we plan to provide an instantiation of $\MEff$ in which also the sequence of operations may be accounted for.
 Since our implementation relies on $\MEff$ we would not have to change
the existing  interpreter and  type checking to accomodate the more refined effects.

We assume operations to be typed. Formally, for each $\opg$, we write $\fun{\opg}{\T_1\ldots\T_n}{\T}$.

The subtyping judgment has shape $\SubT{\T}{\T'}$. Subtyping is the reflexive and transitive closure of the rules of \cref{fig:subtyping}.
\begin{figure}[th]
\begin{math}
\begin{array}{c}
\NamedRule{top}{}{\SubT{\T}{\top}}{}\BigSpace\NamedRule{bot}{}{\SubT{\bot}{\T}}{}\\[3ex]
\NamedRule{sub-fun}
{\SubT{\T'_1}{\T_1}\BigSpace\SubT{\T_2}{\T_2'}}
{\SubT{\FunType{\T_1}{\eff}{\T_2}}{\FunType{\T_1'}{\eff'}{\T'_2}}}
{\SubE{\eff}{\eff'}}
\end{array}
\end{math}
\caption{Subtyping}\label{fig:subtyping}
\end{figure}
In \refToRule{sub-fun} inclusion of effect types is propagated to functional types. So a function producing fewer effects can be used where one producing more effects is needed. Moreover subtyping is, as expected, covariant/contravariant on the result/parameter of functions. 

\begin{figure}[th]
\begin{math}
\begin{array}{l}
\\[-1ex]
\NamedRule{t-var}{}
{ \IsWFExp{\Gamma}{\x}{\T} }
{\Gamma(\x)=\T } 
\quad
\NamedRule{t-abs}{
  \IsEffExp{\Gamma,\TVar{\T}{\x}}{\e}{\T'}{\eff}
}
{ \IsWFExp{\Gamma}{\Fun{\x{:}\T}{\e}}{\FunType{\T}{\eff}{\T'}} }
{ }
\\[3ex]
\NamedRule{t-rec}{
  \IsEffExp{\Gamma,\TVar{\FunType{\T}{\eff}{\T'}}{\f},\TVar{\T}{\x}}{\e}{\T''}{\eff'}
}
{ \IsWFExp{\Gamma}{\RecFun{\f{:}\FunType{\T}{\eff}{\T'}}{\x}{\e}}{\FunType{\T}{\eff}{\T'}} }
{ \SubTE{\T''}{\T'}\wedge\SubE{\eff'}{\eff} }
\\[3ex]
\end{array}
\end{math}

\hrule

\begin{math}
\begin{array}{l}
\\
\NamedRule{t-app}{
  \begin{array}{l}
  \IsWFExp{\Gamma}{\ve_1}{\FunType{\T_1}{\eff}{\T}}\\
  \IsWFExp{\Gamma}{\ve_2}{\T_2}
  \end{array}}
  {  \IsEffExp{\Gamma}{\App{\ve_1}{\ve_2}}{\T}{\eff}}
{\SubT{\T_2}{\T_1}}
\\[5ex]
\NamedRule{t-cond}{ 
  \begin{array}{l}
  \IsWFExp{\Gamma}{\ve}{\Bool}\\
 \IsEffExp{\Gamma}{{\e_1}}{\T_1}{\eff_1}\\ 
 \IsEffExp{\Gamma}{{\e_2}}{\T_2}{\eff_2} 
  \end{array}}
  {  \IsEffExp{\Gamma}{ \If{\ve}{\e_1}{\e_2} }{\T_1\join\T_2}{\eff_1\join\eff_2}}
{}
\\[5ex]
\NamedRule{t-op}{\IsWFExp{\Gamma}{\ve_i}{\T'_i}\ \ \forall i\in 1..n}{\IsEffExp{\Gamma}{\opg(\seq\ve)}{\T}{ \ebasic ( \opg ) }}
{
\seq\ve=\ve_1,\ldots,\ve_n\\
\fun{\opg}{\T_1\ldots\T_n}{\T}\\
\SubT{\T'_i}{\T_i} \ \forall i\in 1..n
}
\\[5ex]
\NamedRule{t-ret}{
  \IsWFExp{\Gamma}{\ve}{\T}
  }
{ \IsEffExp{\Gamma}{\Ret\ve}{\T}{\eun} }
{ } 
\\[5ex]
\NamedRule{t-do}{
  \IsEffExp{\Gamma}{{\e_1}}{\T_1}{\eff_1} 
  \BigSpace
  \IsEffExp{\Gamma,\TVar{\T_2}{\x}}{{\e_2}}{\T}{\eff_2} 
}
{ \IsEffExp{\Gamma}{\Do\x{\e_1}{\e_2}}{\T}{\EComp{\eff_1}{\eff_2}} }
{ \SubT{\T_1}{\T_2}  } 
\end{array}
\end{math}
\caption{Type-and-effect system}\label{fig:typing}
\end{figure}


The typing judgment for values, top part of \cref{fig:typing}, has shape $\IsWFExp{\Gamma}{\ve}{\T}$, since value expressions have no effects. The one for expressions, bottom part of \Cref{fig:typing,fig:typing-handlers}, has shape $\IsEffExp{\Gamma}{\e}{\T}{\eff }$, where $\eff$ is an over-approximation of the effects raised by the evaluation of the expression.

Considering the rules for value expressions, note that the variable of abstractions is decorated with a type, rule \refToRule{t-abs}, and so is the variable denoting the function in recursive definitions,  rule \refToRule{t-rec}. This is due to the need to make type checking algorithmic. To this aim we also added explicit subtyping and subeffecting as a side condition of the rules, instead of having a more general subsumption rule.

The rules for expressions are standard. To type conditional expressions we define the {\em join of types}, $\T_1\join\T_2$ by:
\begin{quoting}
$
\T_1\join\T_2=\begin{cases}
\T_2 &\text{if }\T_1\leq\T_2\\
\T_1 &\text{if }\T_2\leq\T_1\\
\FunType{\T'_1\meet\T'_2}{\eff_1\join\eff_2}{\T''_1\join\T''_2} &\text{if }\T_i=\FunType{\T'_i}{\eff_i}{\T''_i}\\
\top &\text{otherwise }
\end{cases}
$
\end{quoting}
The {\em meet of types}, $\T_1\meet\T_2$ is defined symmetrically.  
The effect of a conditional expression, rule \refToRule{t-cond}, is the join of the effects of the expressions of its branches, whereas for the \kw{do} construct, rule \refToRule{t-do}, it is the monoid composition of effect types.

In \cref{fig:typing-handlers} we show the typing rules for expressions with handlers.  In the rules $\hfilter=\HFilter{\seq\cfilter}{\eff}$ denotes a \emph{(handler) filter}, where
$\cfilter=\CFilter{\opg}{\mode}{\eff}$ is a \emph{(clause) filter}. Given a handler  $\handler=\Handler{\seq\hc}{\x}{\e}$, clause filters record the effects of the expression handling the operation $\opg$ in $\seq\hc$, and handler filters collect the clause filters for all the clauses $\seq\hc$ of  $h$ and handler and and $\eff$ is the effect of the continuing expression $\e$. 

\begin{figure}[th]
\begin{math}
\begin{array}{l}
\NamedRule{t-with}{
\begin{array}{l}
\IsEffExp{\Gamma}{\e}{\T}{\eff}\\
\IsWFHandler{\Gamma}{\T'}{\handler}{\T''}{\hfilter}
\end{array}
}{\IsEffExp{\Gamma}{\With{\handler}{\e}}{\T''}{\FilterFun{\eff}{\hfilter}}}
{
\SubT{\T}{\T'} 
} 
\\[3ex]
\NamedRule{t-handl}
{\begin{array}{l}
\IsEffExp{\Gamma,\TVar{\T}{\x}}{\e'}{\T'}{\eff'}\\
\IsWFClause{\Gamma}{\T''}{\hc_i}{\cfilter_i}
\end{array}
}{\IsWFHandler{\Gamma}{\T}{\Handler{\hc_1\ldots\hc_n}{\x}{\e'}}{\T''}{\HFilter{\cfilter_1\ldots\cfilter_n}{\eff'}}
}{
\SubT{\T'}{\T''} 
}
\\[3ex]
\NamedRule{t-cont}{
  \IsEffExp{\Gamma,\TVars{\T}{\x}} {\e} {\T''} {\eff'}
}{\IsWFClause{\Gamma}{\T'}{\HC{\opg}{\seq\x}{\e}{\Continue}}{\CFilter{\opg}{\Continue}{\eff'}}}
{
  \fun{\opg}{\seq\T}{\T} \\ 
  \SubT{\T''}{\T}  
}
\\[3ex]
\NamedRule{t-stop}{
  \IsEffExp{\Gamma,\TVars{\T}{\x}} {\e} {\T''} {\eff'}
}{\IsWFClause{\Gamma}{\T'}{\HC{\opg}{\seq\x}{\e}{\Stop}}{\CFilter{\opg}{\Stop}{\eff'}}}
{
  \fun{\opg}{\seq\T}{\T} \\ 
  \SubT{\T''}{\T'} 
}
\end{array}
\end{math}
\caption{Typing rules for handlers}\label{fig:typing-handlers}
\end{figure}
In rule \refToRule{t-with}, to typecheck an expression with a handler, we first figure out the type and effect of the expression being handled. We use the type to typecheck the handler itself, specifically, as (a subtype of) the type of the parameter of the final expression, as described in rule \refToRule{t-handl}.

For each clause, in rules \refToRule{t-cont} and \refToRule{t-stop},  we extract the associated filter  and check that the type returned by the expression associated to the clause be compatible with
the final result of the handler. For $\Stop$-clauses the type must be compatible with the one of the final expression of the handler ($\T'$ of  \refToRule{t-handl}).
 A $\Continue$-clause is meant to provide alternative code to be executed before the final expression, hence the type of the clause expression should be (a subtype of) the 
 return type of the operation.

The filter $\FilterFun{\eff}{\HFilter{\cfilter_1\ldots\cfilter_n}{\eff'}}$ produces an effect which takes into consideration the fact that some operations in $\eff$ may be caught and,
moreover, the execution of the catcher may produce some effects. The definition is as follows:
\begin{quoting}
\begin{math}
(1)
\begin{array}{l}
\FilterFun{\emptyset}{\HFilter{\seq\cfilter}{\eff'}}=\eff'
\\
\FilterFun{\{\opg\}{\cup}\eff}{\HFilter{\seq\cfilter}{\eff'}}=\FilterFun{\eff}{\HFilter{\seq\cfilter}{\eff'}}{\cup}\begin{cases}
    \eff'' & \text{if }\CFilter{\opg}{\_}{\eff''}{\in}\seq\cfilter\\
    \{\opg\}  & \text{otherwise}
\end{cases}
\end{array}
\end{math}  
\end{quoting}
We consider each operation $\opg\in\eff$: if there is a clause catching $\opg$, we add the effect of the associated catcher-expression to the filter, otherwise we add the operation $\opg$, since this is not handled. The effect of the continuation is added to the filter.

We can now define the sets 
$\VWT[\ty]$ and $\EWT[\ty,\ef]$ by
\begin{itemize}
\item $\VWT[\T](\ve)$ iff $\IsWFGroundExp{\ve}{\T'}$ for some $\T'$ such that $\SubT{\T'}{\T}$, and 
\item $\EWT[\T,\eff](\e)$ iff $\IsEffGroundExp{\e}{\T'}{\eff'}$ for some $\T',\eff'$ such that $\SubT{{\T'}}{\eff}$ and 
$\SubEI{{\eff'}}{\eff}$.
\end{itemize}


\section{The Implementation}
\label{sec:implementation}

The implementation is done in Haskell \cite{haskell2010}, chosen for its functional paradigm, powerful and
expressive type system, and built-in support for monads.

The implementation adopts a modular architecture in which the reduction framework is completely
independent of the actual language and effects; in other words, the reductions remain operational
regardless of modifications to the language or the introduction of new computational effects.

\subsection{Modules}
\label{sec:modules}

\paragraph{Reduction Interface}
This module corresponds to the abstract framework for monadic operational semantics. 
It is parametric over the language's set of values $\Val$, the set of expressions $\Exp$ and a monad $M$,
and it includes:
\begin{itemize}
    \item The monadic one-step reduction $ \to \subseteq \Exp \times M \Exp $ discussed in \cref{def:mnd-sem}.
    
    \item The definition of the results, $ \Res = \Val + \Wrng $, which model successful terminations or stuck computations.
    \begin{lstlisting}
    data Res v = Ok v | Wr
    \end{lstlisting}
    \item The configurations, $ \Conf = \Exp + \Res $, that make the reduction (of the pure language) $ \to $ a total function.
    \begin{lstlisting}
    data Conf e v = ExpConf e | ResConf (Res v)
    \end{lstlisting}
    \item The monadic reduction $ \longrightarrow_{\text{step}} \subseteq \Conf \times M \Conf $ from configurations to monadic 
      configurations.
    \begin{lstlisting}
    reduceStep::(Reducible m e v)
        => Conf e v -> m (Conf e v)
    \end{lstlisting}
    \item The Kleisli extension $ \Rightarrow \subseteq M \Conf \times M \Conf $ which reduces monadic
      configurations.
    \begin{lstlisting}
    reduceMnStep::(Reducible m e v)
        => m (Conf e v) -> m (Conf e v)
    \end{lstlisting}
    \item The finitary semantics $\finsem{e}$ which describes the monadic result
      of a computation terminating in a finite number of reduction steps.
    \begin{lstlisting}
    evalFin::(Reducible m e v) => e -> m (Res v)
    \end{lstlisting}
\end{itemize}

Any language $ \lang = \langle \Exp, \Val, \valtoexp \rangle $, as defined in \cref{def:mnd-sem}, can be equipped with a monadic operational semantics by implementing this module's interface.
More specifically, an instance of the following type classes must be provided.
\begin{lstlisting}
class Valuable e v | e -> v where
    toVal::e -> Maybe v
    
class(Monad m,Valuable e v)=>Reducible m e v where
    reduce::e -> Maybe (m e)
\end{lstlisting}

The function \lstinline{toVal} encodes the $\valtoexp$ injection, while \lstinline{reduce} represents the monadic one-step reduction, using the \lstinline{Maybe} type constructor to model partiality.

Note that the functional dependency \lstinline{e -> v} is required to guarantee that the value type is uniquely determined by the expression type, allowing the type checker to infer \lstinline{v} unambiguously.

\paragraph{Language specification}

In this module, the Haskell implementation of the extended $\lambdaEff$ abstract syntax is provided, alongside the corresponding operational semantics. 
In \cref{fig:syntax-haskell} we show this mapping, referring to \cref{fig:syntax}.

\begin{figure}[th]
\begin{lstlisting}
    data Val sig e where
        NatVal::(Nat n, Show n) => n -> Val sig e
        RecLamVal ::
            Id -> ArrType' e -> Id ->
            Exp sig e -> Val sig e
        IdVal::Id -> Val sig e
        | ...

    data Exp sig e
        = App (Val sig e) (Val sig e)
        | Ret (Val sig e)
        | Do Identifier (Exp sig e) (Exp sig e)
        | Magic sig [Val sig e]
        | HandleWith (Exp sig e) (Handler sig e)
        | ...

    data Handler sig e = Handler
    { handlerClauses::Map sig (Clause sig e)
    , handlerFinal::(Identifier, Exp sig e)
    }

    data Clause sig e = Clause
    { clauseMode::Mode
    , clauseParams::[Identifier]
    , clauseBody::Exp sig e
    }
\end{lstlisting}
\caption{$\lambdaEff$ syntax in Haskell}\label{fig:syntax-haskell}
\end{figure}

Type constructors such as \lstinline{Val} and \lstinline{Exp}, corresponding respectively to $v$ and $e$ in \cref{fig:syntax}, are generic over the parameters \lstinline{sig} and \lstinline{e}. The former abstracts the syntax over the operation signature, while the latter allows the framework to support different effect algebras.

It is worth focusing on the \lstinline{Magic} data constructor, which represents operation calls. More specifically, it encapsulates an operation identifier and a list of arguments, essentially encoding the syntactic occurrence of an algebraic operation.

To define the semantics of operations in a monadic context, the type class \lstinline{MonSem} is used:
\begin{lstlisting}
class (Monad m, Sig sig) => MonSem m sig where
    run::sig -> [Val sig e] -> m (Val sig e)
\end{lstlisting}

When providing an instance of \lstinline{MonSem}, a family of operations $\Sigma$ that implements \lstinline{Sig} defines the already-presented function ${\pfun{\mrun{\opg}}{\Val^k}{\mfun\Val}}$ in the context of the monad that instantiates the parameter \lstinline{m}.

To complete the language, the \lstinline{Valuable} type class is instantiated by \lstinline{Exp} and \lstinline{Val}:
\begin{lstlisting}
  instance Valuable (Exp sig e) (Val sig e) where
        toVal e = case e of
            Ret v -> Just v
            _ -> Nothing
\end{lstlisting}
As a consequence, a generic monadic operational semantics is provided for our language, effectively bridging pure reductions with those involving effects such as \lstinline{Magic} expressions:
\begin{lstlisting}
  instance (MonSem m sig, Ord sig)
      => Reducible m (Exp sig e) (Val sig e) where
      reduce e = ...
\end{lstlisting}

\paragraph{Type-and-effect algebra}

This module implements the elements needed to define $\lambdaEff$'s static semantics.
Although it is a stand-alone component in the dependency hierarchy, it is logically tied to the specifics of our language.
It defines both value types $\T$ and expression 
types ${\TEff{\T}{\eff}}$. The relevant data types are shown in \cref{fig:types-haskell}.

\begin{figure}[th]
\begin{lstlisting}
newtype ArrType' e=ArrType'(ValType e,e,ValType e)
    
data ValType e
      = NatType    
        | BoolType
        | ArrType (ArrType' e)
        | BotType
        | TopType

    newtype ExpType e = ExpType ((ValType e), e)
\end{lstlisting}
\caption{Value types and expression types in Haskell}\label{fig:types-haskell}
\end{figure}

As stated already, the data structure for effect types is generic over a lattice-ordered monoid
to allow effect composition and sub-effecting.
Furthermore, the module includes the necessary structures to support handlers by implementing handler filters $\hfilter$ and their associated function
$\FilterFEx{\hfilter}: \EffSet \to \EffSet$, which can be applied to an effect type $E$ to derive the filtered transformation. In \cref{fig:effects-handlers-static} we report how these concepts are implemented.

\begin{figure}
    \begin{lstlisting}
class (Lattice e, Monoid e) => Effect e sig where
   applyFilter::e -> Filter sig e -> e
            ...

data Filter sig e = Filter {
   filterClauses::Map sig (ClauseFilter sig e)
     , filterEffect::e
}    
    \end{lstlisting}
    \caption{Effects and handlers structure} 
    \label{fig:effects-handlers-static}
\end{figure}

It is worth reiterating that both types and effects must be compatible with the \lstinline{Lattice} type class, which we formalize as follows:
\begin{lstlisting}
    class (PartialOrd a) => Lattice a where
        join::a -> a -> a
        meet::a -> a -> a
\end{lstlisting}

The \lstinline{EffectSet} type constructor encodes the set-based effect approximation, which simply represents effects as sets of operations:
\begin{lstlisting}
    newtype EffectSet s = EffectSet (Set s)
\end{lstlisting}
This type is equipped with the necessary algebraic structure by instantiating the previously discussed type classes.

\paragraph{Type-and-effect checker}

This module implements the static semantics of $ \lambdaEff$ seen in \cref{fig:typechecking}. It depends on both the previously defined type-and-effect algebra and the language syntax.

It is responsible for verifying the well-typedness of an expression $e$, meaning that given a type context $\Gamma=\overline{x:T}$, associating variables to types, the type-and-effect judgment $\IsWFExp\Gamma\e{\TEff{\T}{\eff}}$ is 
derivable from the rules of the type system. In this case:
\begin{itemize}
    \item $T$ is the type of the value $e$ evaluates to (if it terminates).
    \item $E$ is an upper-bound on the effects that could be raised from the evaluation of  $\e$.
\end{itemize}

As a requirement for some typing rules, the checker enforces subtyping and subeffecting constraints,
and computes the effect transformation for expressions with handlers.

The public interface for type checking is provided by the \lstinline{typeOf} function, which wraps the function \lstinline{typeOfExp}. This latter function recursively traverses the expression tree to implement the inductive structure of the typing rules, building upon (through mutual recursion) the functions \lstinline{typeOfVal}, which infers types for \lstinline{Val} instances and  \lstinline{typeOfClause}, used for application of the rule \refToRule{t-with} for expressions with handlers.

\begin{figure}[th]
\begin{lstlisting}
typeOfVal::(Sig sig, Ord sig, Effect e sig)
  => Val sig e -> Context e ->
  Failable (ValType e) e
        
typeOfClause::(Sig sig, Ord sig, Effect e sig) =>
  sig -> Clause sig e -> Context e ->
  Failable (ExpType e) e
        
typeOfExp::(Sig sig, Ord sig, Effect e sig) =>
  Exp sig e -> Context e -> Failable (ExpType e) e
        
typeOf::(Sig sig, Ord sig, Effect e sig) =>
  (Exp sig e) -> Failable (ExpType e) e
\end{lstlisting}
\caption{Signatures of type-and-effect checking functions}
\label{fig:typechecking}
\end{figure}

Finally, as listed in \cref{fig:typechecking}, the algorithm is generic over the parameters \lstinline{sig} and \lstinline{e} by leveraging the abstract interfaces provided by the type classes \lstinline{Sig}, \lstinline{Lattice} and \lstinline{Effect}. 

\paragraph{Effects Modules}

By definition, the $\lambdaEff$ calculus is generic over $\Sigma$, a family of sets $\{\Sigma_k\}_{k\in\N}$ of $k$-ary operations  raising effects. Each of these modules represents a possible instantiation of $ \Sigma $ and is defined by its static interface and its monadic interpretation. In \cref{fig:ndsighaskell} we provide the formalization of exceptions and non-determinism discussed in  \Cref{ex:exception-mnd,ex:pow-mnd,ex:genericEffect}.

\begin{figure}[th]
    \begin{lstlisting}
data ExceptionSig e = Raise e

instance Sig (ExceptionSig e) where
  arity (Raise _) = ([], BotType)

instance MonSem (Either e) (ExceptionSig e) where
  run (Raise e) [] = Left e
  run _ _ = undefined
    \end{lstlisting}
  \vspace{1em}
    \hrule
    \vspace{1em}
    \begin{lstlisting}
data NDSig = Choose deriving (Show, Eq, Ord)
        
instance Sig NDSig where arity Choose=([],BoolType)
        
instance MonSem [] NDSig where
  run Choose [] = [BoolVal True, BoolVal False] 
  run Choose _ = undefined
    \end{lstlisting}
    \caption{The exception and nondeterminism effects}
    \label{fig:ndsighaskell}
\end{figure}

\begin{description}
    \item[Static signature] This aspect models the family of operations $\Sigma$. When implementing it, a module defines the set of available operation names through a data type (e.g., \lstinline{ExceptionSig} and \lstinline{NDSig}) and specifies their arities by instantiating the \lstinline{Sig} type class.
    
    \item[Monadic interpretation] This component connects the language syntax to a monad $M$ (e.g. \lstinline{Either e}). It implements the partial function $\mrun{\opg} : \Val^k \rightharpoonup M \Val $ required by the operational semantics. In other words, this dictates how the syntactic operation $\mathit{op}$ produces a monadic value, providing the means to raise effects that the monad structure lacks. In our implementation, this is achieved by providing a \lstinline{MonSem} instance.
\end{description}

\section{Examples}\label{sec:examples}
To practically illustrate the monadic reduction, we discuss the following examples. The first one models nondeterminism using the list monad, as described in \cref{ex:pow-mnd}, to produce the output vector of the truth table for a boolean function. The second example showcases the semantics of handlers, defined in \cref{fig:pure-red-handlers}, in the context of exceptions (see \cref{ex:exception-mnd}).

Extending the grammar defined in \cref{fig:syntax}, we force the program to be enclosed in a block that states the chosen signature and monad.

\subsection{Nondeterminism}
\begin{figure}[th]
\textbf{Source code}
    \begin{lstlisting}
using Nondeterminism @ List {
  do xor = return lambda a: Bool.
    return lambda b: Bool.
      if a then 
        if b then return false else return true
      else 
        return b
  ; 
  do x = choose(); do y = choose(); do z = choose();
    
  do x_ = xor x; do x_y = x_ y; do x_y_ = xor x_y;
  x_y_ z
}
    \end{lstlisting}

    \vspace{1em}
    \hrule
    \vspace{1em}

    \textbf{Reduction trace}
    \begin{lstlisting}
    > [do x = choose(); ...]
    > [ do y = choose(); {x = T} ...,
        do y = choose(); {x = F} ... ]
    > [ do z = choose(); {x = T, y = T} ..., 
       ...,
       do z = choose(); {x = F, y = F} ... ]
    ...
    > [ x_y_ T {x_y = F, z = T} ..., 
        ..., 
        x_y_ F {x_y = F, z = F} ... ]
    > [ T, F, F, T, F, T, T, F ]
    \end{lstlisting}
    \caption{Nondeterministic XOR under list monad}
    \label{fig:ndlist}
\end{figure}

In the example of \cref{fig:ndlist}, the boolean function is the exclusive OR ($a \oplus b$), defined through curried lambda abstractions and nested conditional expressions. The objective is to compute the ordered sequence of all possible outcomes for the expression $x \oplus y \oplus z$, where $x, y, z \in \mathbb{B}$.
We achieve this by first nondeterministically binding the result of the \lstinline{choose()} operation to the variables \lstinline{x},  \lstinline{y}, and  \lstinline{z}. Then, the instantiated result is obtained via a sequence of intermediate bindings and  \lstinline{xor} function applications, essentially following the structure  \lstinline{xor ((xor x) y) z}.

The reduction trace highlights the state-space expansion caused by the list monad. After reducing the third occurrence of \lstinline{choose()} , the computation contains all eight needed configurations, each of which calculates the result for the corresponding assignment. It is worth pointing out that the final output follows an inverse order compared to standard truth table conventions. This is a consequence of the semantics provided in \cref{ex:genericEffect:nd} of \cref{ex:genericEffect} where $\mrun{\chooseop}$ evaluates to $[\True,\False]$ rather than $[\False,\True]$.

\subsection{Exceptions with handlers}
\begin{figure}
    \begin{lstlisting}
using Exceptions @ Exceptions {
  handle
    do safe_pred = return lambda a: Nat.
      do t = iszero a; if t then
        raise<underflow>()
      else
        pred a
    ;
    do safe_div = return lambda a: Nat.
      return lambda b: Nat.
        do t = iszero b; if t then
          raise<div_zero>()
        else
          a / b
    ;
    do f = return lambda a: Nat.
      return lambda b: Nat.
        do a_ = safe_div a; do a_b = a_ b;
        safe_pred a_b
    ; 
    do a_ = f ...;
    a_ ...
  with {
    raise<underflow>() ->c raise<math_error>(),
    raise<div_zero>() ->c raise<math_error>(),
    x -> return x
  }
}

    \end{lstlisting}
    \vspace{1em}
    \hrule
    \vspace{1em}
    \textbf{Reduction trace (\lstinline{f 0 1})}
    \begin{lstlisting}
    > R(handle (do a_ = f 0; a_ 1) with h)
    > R(handle (
         do a_b = return 0; safe_pred a_b) with h)
    > R(handle (safe_pred 0) with h )
    > R(handle (raise<underflow>()) with h)
    > R(do x = raise<math_error>(); return x)
    > L(math_error)
    \end{lstlisting}

    \vspace{1em}

    \textbf{Reduction trace (\lstinline{f 1 0})}
    \begin{lstlisting}
    > R(handle (do a_ = f 1; a_ 0) with h)
    > R(handle (
        do a_ =safe_div 1; do a_b = a_ 0;..)with h)
    > R(handle (raise<div_zero>()) with h)
    > R(do a_b=raise<math_error>();(safe_pred a_b))
    > L(math_error )
    \end{lstlisting}
    \caption{Exceptions unification via handler}
    \label{fig:exchandler}
\end{figure}

The second example, illustrated in \cref{fig:exchandler}, demonstrates the usage of handlers for exception unification to obtain error abstraction. This pattern is commonly applied whenever the internal logic details do not need to be exposed.
 Given two natural numbers $a, b$, note that the expression $(a /b) - 1$ is undefined in $\mathbb{N}$ when either $a = 0$ or $b = 0$. Our attempt is to classify these cases as ``generic math errors''. 
 To do this, we define the functions \lstinline{safe\_pred} and \lstinline{safe\_div}. The function \lstinline{safe\_pred a} evaluates to \lstinline{a-1} and raises a domain violation (``underflow'') exception when \lstinline{a = 0}.  
The function \lstinline{safe\_div a b} evaluates to \lstinline{a / b} and raises a division by zero (``div\_zero'') exception when \lstinline{b = 0}. Finally, we define \lstinline{f a b}, which evaluates to the final expression. 

Note that in our implementation, the \lstinline{Exceptions} monad is mapped to the underlying Haskell \lstinline{Either String} monad. We use \lstinline{L} and \lstinline{R} as shorthands for \lstinline{Left} and \lstinline{Right}, respectively.
The reduction traces show how, in both scenarios, the specific exceptions are first raised by the corresponding inner function and later intercepted and abstracted by the handler's clauses with \textit{continue} mode, which replaces the raised exception with the more general one.
\section{Related Work}\label{sec:related}
 To better position our work within the current landscape, we examine several approaches to algebraic effects adopted by different systems. 

{\bf Koka}~\cite{koka-lang} is a call-by-value functional language that unifies various control-flow constructs, such as exceptions, iterators, and async-await, via algebraic effects.
Its type system, which supports automatic type-inference, is characterized by row polymorphism with scoped labels, which precisely tracks effects in a concise and simple way. Operationally, handlers and the \textit{resume} construct treat effects as delimited continuations, allowing a program to be suspended and later resumed at the point where the operation was invoked. To achieve high performance on call-stack-abstracted architectures, the compiler employs a type-directed continuation passing style (CPS) translation to exclusively target effectful program portions, complemented by polymorphic duplication.

{\bf Frank}~\cite{LindleyMM16} is a functional language that distinguishes between values (being) and computations (doing). Here, algebraic effects are called abilities and are collections of interfaces; each interface describes the operators it supports. The bidirectional type system does not support effect inference and employs silent effect polymorphism, which eliminates the need for explicit effect variables.
A key feature is the absence of explicit handlers: functions are generalized into n-ary operators (or multihandlers). Rather than using a dedicated construct, effects are handled by application: an operator acts as a handler if its input ports include adjustments, allowing it to coordinate multiple computations simultaneously. Operationally, the continuations in Frank are shallow: the handler is not automatically re-installed upon resumption, and explicit recursion is required for deep-handling.

{\bf Effekt}~\cite{effekt-lang} is a Scala library that takes a different approach by embedding effects directly into the host language via a capability-passing style. Effects are Scala objects treated as capabilities and are passed as arguments to functions. To perform an effect, the program simply calls a method defined on the corresponding capability object, following a standard object-oriented fashion.
To ensure safety, capabilities must be guaranteed not to be used after the handler has finished. Effekt achieves this by leveraging Scala's intersection types and path-dependent types, enabling support for subtyping and polymorphism ``for free'', which are built-in in Scala.
Operationally, the library runs on a monad for multi-prompt delimited continuations, where possibly-nested handlers serve the dual purpose of introducing capabilities and delimiting the scope for the continuation capture.

{\bf Links}~\cite{links-lang} is a tierless web-oriented language based on a call-by-value calculus which, similarly to Koka, uses row-polymorphism to track effects, but it adopts a stricter Remy-style approach where the absence of an effect is made explicit. It is characterized by the use of a higher-order CPS complete translation from a formalism with effects and handlers to a formalism with neither. This transformation is tail-recursive and allows for the elimination of administrative redexes at compile time, properties that are essential given its Javascript backend's limitations. The handler semantics is originally defined in a deep way, but alternative rules for shallow handlers are also provided.

{\bf Helium}~\cite{Biernacki19} is an experimental language based on the $\lambda^{HEL}$ calculus, addressing the problem of hiding and isolating algebraic effects. Unlike other systems where effect operations are always globally visible, Helium introduces existential effects to treat them as ``black boxes'' that conceal the implementation details from the user. It also provides local effects to ensure that internal operations cannot be ``stolen'', i.e. intercepted by external handlers. To match the operations to the correct handlers, Helium implements effect coercions, which act as explicit markers to guide operations through the handlers stack. These features are organized in an ML-style module system, and a specific CEK-like abstract machine is described as the execution model.

{\bf Handlers in Action}~\cite{Kammar2013} introduces an implementation of algebraic effects as a Haskell library, presented as a modular alternative to the inherently rigid monad transformer stacks. The authors leverage the Template Haskell and quasiquote features of GHC to declare operations and handlers using a clean, declarative domain-specific syntax.
Modular instantiation is achieved via open handlers, which interpret a targeted subset of operations while forwarding the rest to the outer environment, enabling flexible composition.
The framework supports both deep and shallow handlers and is backed by the $\lambda_{eff}$ calculus, for which termination and soundness proofs are provided.

{\bf Multicore OCaml}~\cite{Sivaramakrishnan2021}, which has now been merged into OCaml, introduces algebraic effects as a foundation for direct-style concurrency while maintaining complete backwards compatibility. Effects are treated as extensible variant types, without enforcing static effect-safety. Handler execution is made possible via heap-allocated fibers, which provide a stackful context for one-shot continuations. The implementation is shown to be efficient for new code utilizing handlers while imposing minimal overhead on existing programs.

\paragraph{Comparison}
All of these systems differ from what we propose in two main aspects:
\begin{itemize}
    \item \textbf{Continuations.} Unlike the cited projects, $\lambdaEff$ is based on generic effects and has no support for explicit continuations. Our non-standard handlers provide a more constrained mechanism via their fixed modes \textit{continue} and \textit{stop}, rather than an explicit \textit{resume} construct.
    \item \textbf{Monadic semantics.} Our approach is characterized by its monadic operational semantics, where small-step reduction is parametric over the underlying monad. In other algebraic effect systems, operations derive semantics exclusively through handler interpretation. In contrast, we associate each operation with a specific monadic action, while still allowing for handler interception. 
\end{itemize}
\section{Conclusions}
\label{sec:conclusion}

In this paper we presented an implementation in Haskell of a framework for defining a small-step semantics and a type system for languages with generic effects introduced in~\citet{DGZ25}. The framework is an instantiation of $\lambdaEff$, a functional call-by-value lambda calculus with operations that raise effects and effect handlers. The implementation, whose code can be found at~\cite{Raviola25},  provides a small-step interpreter, which can be  iterated to produce an evaluator, and a type-and-effect-checker for $\lambdaEff$. The choice of Haskell was due to its powerful and expressive type system, and built-in support for monads. The project shows that the theoretical framework of~\cite{DGZ25} can be translated into a concrete, modular, executable system. 

The system represents a foundational prototype rather than a complete language environment. It has a front end in which the user can specify an expression and the target monad. Then the step-by-step reduction is  executed after checking the correctness of the program. 

As future work, we plan to increase the expressiveness of both the language and the effect system.

On the language side, we intend to introduce additional constructs and to allow users to define their own monads by specifying the required data types and effectful operations. To support this, we are exploring the meta-programming capabilities of Template Haskell, as described in~\cite{SPJ02}, to automatically generate the necessary \lstinline{MonSem} instance described at the end of~\cref{sec:implementation}.

On the effect-type side, our current implementation represents effect types as sets of operations. In contrast,~\cite{DGZ25} models them as sets of (possibly infinite) sequences of operations. As a result, the implementation is less expressive. In particular, it cannot approximate computational effects in which the order of operations matters, for example, considering an output monad, sequences of write operations.

To address this limitation, we are investigating a formalism based on regular expressions that can effectively describe possibly infinite sequences of operations. This would allow us to approximate the (possibly infinite) effects of recursive functions.
\begin{acks}
We are grateful to the anonymous reviewers for their useful suggestions. This work  has the financial support of the Universit\`a del Piemonte Orientale.
\end{acks}


\begin{thebibliography}{27}


\ifx \showCODEN    \undefined \def \showCODEN     #1{\unskip}     \fi
\ifx \showDOI      \undefined \def \showDOI       #1{#1}\fi
\ifx \showISBNx    \undefined \def \showISBNx     #1{\unskip}     \fi
\ifx \showISBNxiii \undefined \def \showISBNxiii  #1{\unskip}     \fi
\ifx \showISSN     \undefined \def \showISSN      #1{\unskip}     \fi
\ifx \showLCCN     \undefined \def \showLCCN      #1{\unskip}     \fi
\ifx \shownote     \undefined \def \shownote      #1{#1}          \fi
\ifx \showarticletitle \undefined \def \showarticletitle #1{#1}   \fi
\ifx \showURL      \undefined \def \showURL       {\relax}        \fi
\providecommand\bibfield[2]{#2}
\providecommand\bibinfo[2]{#2}
\providecommand\natexlab[1]{#1}
\providecommand\showeprint[2][]{arXiv:#2}

\bibitem[Bauer and Pretnar(2015)]%
        {BauerP15}
\bibfield{author}{\bibinfo{person}{Andrej Bauer} {and} \bibinfo{person}{Matija
  Pretnar}.} \bibinfo{year}{2015}\natexlab{}.
\newblock \showarticletitle{Programming with algebraic effects and handlers}.
\newblock \bibinfo{journal}{\emph{Journal of Logical and Algebraic Methods in
  Programming}} \bibinfo{volume}{84}, \bibinfo{number}{1}
  (\bibinfo{year}{2015}), \bibinfo{pages}{108--123}.
\newblock
\urldef\tempurl%
\url{https://doi.org/10.1016/J.JLAMP.2014.02.001}
\showDOI{\tempurl}


\bibitem[Biernacki et~al\mbox{.}(2019)]%
        {Biernacki19}
\bibfield{author}{\bibinfo{person}{Dariusz Biernacki}, \bibinfo{person}{Maciej
  Pir\'{o}g}, \bibinfo{person}{Piotr Polesiuk}, {and} \bibinfo{person}{Filip
  Sieczkowski}.} \bibinfo{year}{2019}\natexlab{}.
\newblock \showarticletitle{Abstracting algebraic effects}.
\newblock \bibinfo{journal}{\emph{Proc. ACM Program. Lang.}}
  \bibinfo{volume}{3}, \bibinfo{number}{POPL}, Article \bibinfo{articleno}{6}
  (\bibinfo{date}{Jan.} \bibinfo{year}{2019}), \bibinfo{numpages}{28}~pages.
\newblock
\urldef\tempurl%
\url{https://doi.org/10.1145/3290319}
\showDOI{\tempurl}


\bibitem[Brachth{\"a}user et~al\mbox{.}(2026)]%
        {effekt-lang}
\bibfield{author}{\bibinfo{person}{Jonathan Brachth{\"a}user},
  \bibinfo{person}{Philipp Schuster}, {and} \bibinfo{person}{Klaus Ostermann}.}
  \bibinfo{year}{2026}\natexlab{}.
\newblock \bibinfo{title}{{Effekt} Language}.
\newblock \bibinfo{howpublished}{\url{https://effekt-lang.org/}}.
\newblock
\newblock
\shownote{Accessed: 2026-03-01}.


\bibitem[Cooper et~al\mbox{.}(2026)]%
        {links-lang}
\bibfield{author}{\bibinfo{person}{Ezra Cooper}, \bibinfo{person}{Sam Lindley},
  \bibinfo{person}{Philip Wadler}, {and} \bibinfo{person}{Jeremy Yallop}.}
  \bibinfo{year}{2026}\natexlab{}.
\newblock \bibinfo{title}{The {Links} Programming Language}.
\newblock \bibinfo{howpublished}{\url{https://links-lang.org/}}.
\newblock
\newblock
\shownote{Accessed: 2026-03-01}.


\bibitem[Dagnino et~al\mbox{.}(2025)]%
        {DGZ25}
\bibfield{author}{\bibinfo{person}{Francesco Dagnino}, \bibinfo{person}{Paola
  Giannini}, {and} \bibinfo{person}{Elena Zucca}.}
  \bibinfo{year}{2025}\natexlab{}.
\newblock \showarticletitle{Monadic Type-And-Effect Soundness}. In
  \bibinfo{booktitle}{\emph{39th European Conference on Object-Oriented
  Programming, {ECOOP} 2025, Bergen, Norway, June 30 - July 2, 2025}}
  \emph{(\bibinfo{series}{LIPIcs}, Vol.~\bibinfo{volume}{333})},
  \bibfield{editor}{\bibinfo{person}{Jonathan Aldrich} {and}
  \bibinfo{person}{Alexandra Silva}} (Eds.). \bibinfo{publisher}{Schloss
  Dagstuhl - Leibniz-Zentrum f{\"{u}}r Informatik}, \bibinfo{pages}{7:1--7:31}.
\newblock
\urldef\tempurl%
\url{https://doi.org/10.4230/LIPICS.ECOOP.2025.7}
\showDOI{\tempurl}


\bibitem[Eilenberg and Moore(1965)]%
        {EilenbergM65}
\bibfield{author}{\bibinfo{person}{Samuel Eilenberg} {and}
  \bibinfo{person}{John~C. Moore}.} \bibinfo{year}{1965}\natexlab{}.
\newblock \showarticletitle{{Adjoint functors and triples}}.
\newblock \bibinfo{journal}{\emph{Illinois Journal of Mathematics}}
  \bibinfo{volume}{9}, \bibinfo{number}{3} (\bibinfo{year}{1965}),
  \bibinfo{pages}{381 -- 398}.
\newblock
\urldef\tempurl%
\url{https://doi.org/10.1215/ijm/1256068141}
\showDOI{\tempurl}


\bibitem[Jones et~al\mbox{.}(2010)]%
        {haskell2010}
\bibfield{author}{\bibinfo{person}{Simon~Peyton Jones} {et~al\mbox{.}}}
  \bibinfo{year}{2010}\natexlab{}.
\newblock \bibinfo{booktitle}{\emph{Haskell 2010 Language Report}}.
\newblock \bibinfo{type}{Technical Report}. \bibinfo{institution}{The Haskell
  Committee / Haskell.org}.
\newblock
\urldef\tempurl%
\url{https://www.haskell.org/onlinereport/haskell2010/}
\showURL{%
\tempurl}
\newblock
\shownote{Available online}.


\bibitem[Kammar et~al\mbox{.}(2013)]%
        {Kammar2013}
\bibfield{author}{\bibinfo{person}{Ohad Kammar}, \bibinfo{person}{Sam Lindley},
  {and} \bibinfo{person}{Nicolas Oury}.} \bibinfo{year}{2013}\natexlab{}.
\newblock \showarticletitle{Handlers in action}. In
  \bibinfo{booktitle}{\emph{Proceedings of the 18th ACM SIGPLAN International
  Conference on Functional Programming}} (Boston, Massachusetts, USA)
  \emph{(\bibinfo{series}{ICFP '13})}. \bibinfo{publisher}{Association for
  Computing Machinery}, \bibinfo{address}{New York, NY, USA},
  \bibinfo{pages}{145–158}.
\newblock
\showISBNx{9781450323260}
\urldef\tempurl%
\url{https://doi.org/10.1145/2500365.2500590}
\showDOI{\tempurl}


\bibitem[Katsumata(2014)]%
        {Katsumata14}
\bibfield{author}{\bibinfo{person}{Shin{-}ya Katsumata}.}
  \bibinfo{year}{2014}\natexlab{}.
\newblock \showarticletitle{Parametric effect monads and semantics of effect
  systems}. In \bibinfo{booktitle}{\emph{\POPL{41st}{2014}}},
  \bibfield{editor}{\bibinfo{person}{Suresh Jagannathan} {and}
  \bibinfo{person}{Peter Sewell}} (Eds.). \bibinfo{publisher}{{ACM}},
  \bibinfo{pages}{633--646}.
\newblock
\urldef\tempurl%
\url{https://doi.org/10.1145/2535838.2535846}
\showDOI{\tempurl}


\bibitem[Leijen(2026)]%
        {koka-lang}
\bibfield{author}{\bibinfo{person}{Daan Leijen}.}
  \bibinfo{year}{2026}\natexlab{}.
\newblock \bibinfo{title}{The {Koka} Programming Language}.
\newblock
  \bibinfo{howpublished}{\url{https://koka-lang.github.io/koka/doc/index.html}}.
\newblock
\newblock
\shownote{Accessed: 2026-03-01}.


\bibitem[Levy et~al\mbox{.}(2003)]%
        {LevyPT03}
\bibfield{author}{\bibinfo{person}{Paul~Blain Levy}, \bibinfo{person}{John
  Power}, {and} \bibinfo{person}{Hayo Thielecke}.}
  \bibinfo{year}{2003}\natexlab{}.
\newblock \showarticletitle{Modelling environments in call-by-value programming
  languages}.
\newblock \bibinfo{journal}{\emph{Information and Computation}}
  \bibinfo{volume}{185}, \bibinfo{number}{2} (\bibinfo{year}{2003}),
  \bibinfo{pages}{182--210}.
\newblock
\urldef\tempurl%
\url{https://doi.org/10.1016/S0890-5401(03)00088-9}
\showDOI{\tempurl}


\bibitem[Lindley et~al\mbox{.}(2016)]%
        {LindleyMM16}
\bibfield{author}{\bibinfo{person}{Sam Lindley}, \bibinfo{person}{Conor
  McBride}, {and} \bibinfo{person}{Craig McLaughlin}.}
  \bibinfo{year}{2016}\natexlab{}.
\newblock \showarticletitle{Do be do be do}.
\newblock \bibinfo{journal}{\emph{CoRR}}  \bibinfo{volume}{abs/1611.09259}
  (\bibinfo{year}{2016}).
\newblock
\showeprint[arXiv]{1611.09259}
\urldef\tempurl%
\url{http://arxiv.org/abs/1611.09259}
\showURL{%
\tempurl}


\bibitem[Marino and Millstein(2009)]%
        {MarinoM09}
\bibfield{author}{\bibinfo{person}{Daniel Marino} {and}
  \bibinfo{person}{Todd~D. Millstein}.} \bibinfo{year}{2009}\natexlab{}.
\newblock \showarticletitle{A generic type-and-effect system}. In
  \bibinfo{booktitle}{\emph{TLDI'09: Types in Languages Design and
  Implementatio}}, \bibfield{editor}{\bibinfo{person}{Andrew Kennedy} {and}
  \bibinfo{person}{Amal Ahmed}} (Eds.). \bibinfo{publisher}{{ACM} Press},
  \bibinfo{pages}{39--50}.
\newblock
\urldef\tempurl%
\url{https://doi.org/10.1145/1481861.1481868}
\showDOI{\tempurl}


\bibitem[Moggi(1989)]%
        {Moggi89}
\bibfield{author}{\bibinfo{person}{Eugenio Moggi}.}
  \bibinfo{year}{1989}\natexlab{}.
\newblock \showarticletitle{Computational Lambda-Calculus and Monads}. In
  \bibinfo{booktitle}{\emph{\LICS{4th}{1989}}}. \bibinfo{publisher}{{IEEE}
  Computer Society}, \bibinfo{pages}{14--23}.
\newblock
\urldef\tempurl%
\url{https://doi.org/10.1109/LICS.1989.39155}
\showDOI{\tempurl}


\bibitem[Moggi(1991)]%
        {Moggi91}
\bibfield{author}{\bibinfo{person}{Eugenio Moggi}.}
  \bibinfo{year}{1991}\natexlab{}.
\newblock \showarticletitle{Notions of Computation and Monads}.
\newblock \bibinfo{journal}{\emph{Information and Computation}}
  \bibinfo{volume}{93}, \bibinfo{number}{1} (\bibinfo{year}{1991}),
  \bibinfo{pages}{55--92}.
\newblock
\urldef\tempurl%
\url{https://doi.org/10.1016/0890-5401(91)90052-4}
\showDOI{\tempurl}


\bibitem[Nielson and Nielson(1999)]%
        {NielsonN99}
\bibfield{author}{\bibinfo{person}{Flemming Nielson} {and}
  \bibinfo{person}{Hanne~Riis Nielson}.} \bibinfo{year}{1999}\natexlab{}.
\newblock \showarticletitle{Type and Effect Systems}. In
  \bibinfo{booktitle}{\emph{Correct System Design, Recent Insight and
  Advances}} \emph{(\bibinfo{series}{Lecture Notes in Computer Science},
  Vol.~\bibinfo{volume}{1710})},
  \bibfield{editor}{\bibinfo{person}{Ernst{-}R{\"{u}}diger Olderog} {and}
  \bibinfo{person}{Bernhard Steffen}} (Eds.). \bibinfo{publisher}{Springer},
  \bibinfo{pages}{114--136}.
\newblock
\urldef\tempurl%
\url{https://doi.org/10.1007/3-540-48092-7\_6}
\showDOI{\tempurl}


\bibitem[Plotkin and Power(2001)]%
        {PlotkinP01}
\bibfield{author}{\bibinfo{person}{Gordon~D. Plotkin} {and}
  \bibinfo{person}{John Power}.} \bibinfo{year}{2001}\natexlab{}.
\newblock \showarticletitle{Adequacy for Algebraic Effects}. In
  \bibinfo{booktitle}{\emph{\FOSSACS{4th}{2001}}}
  \emph{(\bibinfo{series}{Lecture Notes in Computer Science},
  Vol.~\bibinfo{volume}{2030})}, \bibfield{editor}{\bibinfo{person}{Furio
  Honsell} {and} \bibinfo{person}{Marino Miculan}} (Eds.).
  \bibinfo{publisher}{Springer}, \bibinfo{pages}{1--24}.
\newblock
\urldef\tempurl%
\url{https://doi.org/10.1007/3-540-45315-6\_1}
\showDOI{\tempurl}


\bibitem[Plotkin and Power(2002)]%
        {PlotkinP02}
\bibfield{author}{\bibinfo{person}{Gordon~D. Plotkin} {and}
  \bibinfo{person}{John Power}.} \bibinfo{year}{2002}\natexlab{}.
\newblock \showarticletitle{Notions of Computation Determine Monads}. In
  \bibinfo{booktitle}{\emph{\FOSSACS{5th}{2002}}}
  \emph{(\bibinfo{series}{Lecture Notes in Computer Science},
  Vol.~\bibinfo{volume}{2303})}, \bibfield{editor}{\bibinfo{person}{Mogens
  Nielsen} {and} \bibinfo{person}{Uffe Engberg}} (Eds.).
  \bibinfo{publisher}{Springer}, \bibinfo{pages}{342--356}.
\newblock
\urldef\tempurl%
\url{https://doi.org/10.1007/3-540-45931-6\_24}
\showDOI{\tempurl}


\bibitem[Plotkin and Power(2003)]%
        {PlotkinP03}
\bibfield{author}{\bibinfo{person}{Gordon~D. Plotkin} {and}
  \bibinfo{person}{John Power}.} \bibinfo{year}{2003}\natexlab{}.
\newblock \showarticletitle{Algebraic Operations and Generic Effects}.
\newblock \bibinfo{journal}{\emph{Applied Categorical Structures}}
  \bibinfo{volume}{11}, \bibinfo{number}{1} (\bibinfo{year}{2003}),
  \bibinfo{pages}{69--94}.
\newblock
\urldef\tempurl%
\url{https://doi.org/10.1023/A:1023064908962}
\showDOI{\tempurl}


\bibitem[Plotkin and Pretnar(2009)]%
        {PlotkinPretnar09}
\bibfield{author}{\bibinfo{person}{Gordon~D. Plotkin} {and}
  \bibinfo{person}{Matija Pretnar}.} \bibinfo{year}{2009}\natexlab{}.
\newblock \showarticletitle{Handlers of Algebraic Effects}. In
  \bibinfo{booktitle}{\emph{\ESOP{18th}{2009}}} \emph{(\bibinfo{series}{Lecture
  Notes in Computer Science}, Vol.~\bibinfo{volume}{5502})},
  \bibfield{editor}{\bibinfo{person}{Giuseppe Castagna}} (Ed.).
  \bibinfo{publisher}{Springer}, \bibinfo{pages}{80--94}.
\newblock
\urldef\tempurl%
\url{https://doi.org/10.1007/978-3-642-00590-9\_7}
\showDOI{\tempurl}


\bibitem[Plotkin and Pretnar(2013)]%
        {PlotkinPretnar13}
\bibfield{author}{\bibinfo{person}{Gordon~D. Plotkin} {and}
  \bibinfo{person}{Matija Pretnar}.} \bibinfo{year}{2013}\natexlab{}.
\newblock \showarticletitle{Handling Algebraic Effects}.
\newblock \bibinfo{journal}{\emph{Logical Methods in Computer Science}}
  \bibinfo{volume}{9}, \bibinfo{number}{4} (\bibinfo{year}{2013}).
\newblock
\urldef\tempurl%
\url{https://doi.org/10.2168/LMCS-9(4:23)2013}
\showDOI{\tempurl}


\bibitem[Pretnar(2015)]%
        {Pretnar15}
\bibfield{author}{\bibinfo{person}{Matija Pretnar}.}
  \bibinfo{year}{2015}\natexlab{}.
\newblock \showarticletitle{An Introduction to Algebraic Effects and Handlers.
  Invited tutorial paper}. In \bibinfo{booktitle}{\emph{\MFPS{31st}{2015}}}
  \emph{(\bibinfo{series}{Electronic Notes in Theoretical Computer Science},
  Vol.~\bibinfo{volume}{319})}, \bibfield{editor}{\bibinfo{person}{Dan~R.
  Ghica}} (Ed.). \bibinfo{publisher}{Elsevier}, \bibinfo{pages}{19--35}.
\newblock
\urldef\tempurl%
\url{https://doi.org/10.1016/J.ENTCS.2015.12.003}
\showDOI{\tempurl}


\bibitem[Raviola(2025)]%
        {Raviola25}
\bibfield{author}{\bibinfo{person}{Stefano Raviola}.}
  \bibinfo{year}{2025}\natexlab{}.
\newblock \bibinfo{title}{monadic-semantics-calculus}.
\newblock
  \bibinfo{howpublished}{\url{https://github.com/20051799-uniupo/monadic-semantics-calculus/tree/master}}.
\newblock


\bibitem[Sheard and Jones(2002)]%
        {SPJ02}
\bibfield{author}{\bibinfo{person}{Tim Sheard} {and}
  \bibinfo{person}{Simon~Peyton Jones}.} \bibinfo{year}{2002}\natexlab{}.
\newblock \showarticletitle{Template meta-programming for Haskell}. In
  \bibinfo{booktitle}{\emph{Proceedings of the 2002 {ACM} {SIGPLAN} Workshop on
  Haskell, Haskell 2002, Pittsburgh, Pennsylvania, USA, October 3, 2002}},
  \bibfield{editor}{\bibinfo{person}{Manuel M.~T. Chakravarty}} (Ed.).
  \bibinfo{publisher}{{ACM}}, \bibinfo{pages}{1--16}.
\newblock
\urldef\tempurl%
\url{https://doi.org/10.1145/581690.581691}
\showDOI{\tempurl}


\bibitem[Sivaramakrishnan et~al\mbox{.}(2021)]%
        {Sivaramakrishnan2021}
\bibfield{author}{\bibinfo{person}{KC Sivaramakrishnan},
  \bibinfo{person}{Stephen Dolan}, \bibinfo{person}{Leo White},
  \bibinfo{person}{Tom Kelly}, \bibinfo{person}{Sadiq Jaffer}, {and}
  \bibinfo{person}{Anil Madhavapeddy}.} \bibinfo{year}{2021}\natexlab{}.
\newblock \showarticletitle{Retrofitting effect handlers onto OCaml}. In
  \bibinfo{booktitle}{\emph{Proceedings of the 42nd ACM SIGPLAN International
  Conference on Programming Language Design and Implementation}} (Virtual,
  Canada) \emph{(\bibinfo{series}{PLDI 2021})}. \bibinfo{publisher}{Association
  for Computing Machinery}, \bibinfo{address}{New York, NY, USA},
  \bibinfo{pages}{206–221}.
\newblock
\showISBNx{9781450383912}
\urldef\tempurl%
\url{https://doi.org/10.1145/3453483.3454039}
\showDOI{\tempurl}


\bibitem[Street(1972)]%
        {Street72}
\bibfield{author}{\bibinfo{person}{Ross Street}.}
  \bibinfo{year}{1972}\natexlab{}.
\newblock \showarticletitle{The formal theory of monads}.
\newblock \bibinfo{journal}{\emph{Journal of Pure and Applied Algebra}}
  \bibinfo{volume}{2}, \bibinfo{number}{2} (\bibinfo{year}{1972}),
  \bibinfo{pages}{149 -- 168}.
\newblock
\showISSN{0022-4049}
\urldef\tempurl%
\url{https://doi.org/10.1016/0022-4049(72)90019-9}
\showDOI{\tempurl}


\bibitem[Wright and Felleisen(1994)]%
        {WrightF94}
\bibfield{author}{\bibinfo{person}{Andrew~K. Wright} {and}
  \bibinfo{person}{Matthias Felleisen}.} \bibinfo{year}{1994}\natexlab{}.
\newblock \showarticletitle{A Syntactic Approach to Type Soundness}.
\newblock \bibinfo{journal}{\emph{Information and Computation}}
  \bibinfo{volume}{115}, \bibinfo{number}{1} (\bibinfo{year}{1994}),
  \bibinfo{pages}{38--94}.
\newblock
\urldef\tempurl%
\url{https://doi.org/10.1006/inco.1994.1093}
\showDOI{\tempurl}


\end{thebibliography}


\end{document}